\documentclass[a4paper, 12pt]{article}

\usepackage[dvips]{epsfig}

\usepackage{amsmath}

\usepackage{amssymb}

\begin{document}

\begin{titlepage}

\begin{flushright}

{\sc BI-TP 2004/08}

\end{flushright}

\vspace{40pt}

\begin{center}

{\huge Classical approximation of the Boltzmann equation in high energy QCD}

\end{center}

\vspace{1.5cm}

\begin{center}

$T.~Stockamp^1$

\end{center}

\begin{center}

Fakult\"at f\"ur Physik, Universit\"at Bielefeld, 

D-33501 Bielefeld, Germany

\end{center}

\vspace{1.5cm}





\begin{abstract}

Recently, Mueller and Son discussed the time evolution of a dense system
towards equilibrium in a scalar $\lambda\varphi^4$ field
theory \cite{Mueller:2002gd}. They show the equivalence
of the classical field approximation and the Boltzmann equation in all but
linear terms in the occupation number.
Here we present the generalization to high energy QCD.

\end{abstract}


\vspace{7cm}

$^1mail:$ stockamp@physik.uni-bielefeld.de

\end{titlepage}

\section{Introduction}

In high energy heavy ion collision experiments it is believed that a new 
state of matter, the so-called quark gluon plasma, may be created
(\cite{Ludlam:2003rh}, \cite{Gyulassy:2004zy}). Theoretically it is therefore important to understand 
the time evolution of dense strongly interacting matter towards 
thermal equilibrium.
This evolution can be described in terms of the gluon occupation number
f. In the very early stages after a collision f is of the order 
${\alpha_s}^{-1}$ (\cite{Mueller:1999pi}, \cite{Mueller:1999fp}) where $\alpha_s$ is the strong coupling constant. As 
the system evolves f decreases and is finally of the order 1 at 
equilibrium.   

Unfortunately there is no general theoretical framework to describe the 
complete evolution. For large f (i.e. very early times) non-linear 
classical field theory can be applied while a Boltzmann equation that 
takes quantum fluctuations into account (this will be called full or 
``quantum`` Boltzmann equation) may be used near equilibrium.

In order to obtain a continuous description of the evolution one may try to 
fit the transition between the two approaches by a classical field 
approximation of the full Boltzmann equation (Fig. 1).
\begin{figure}[h]\begin{center}
\epsfig{file=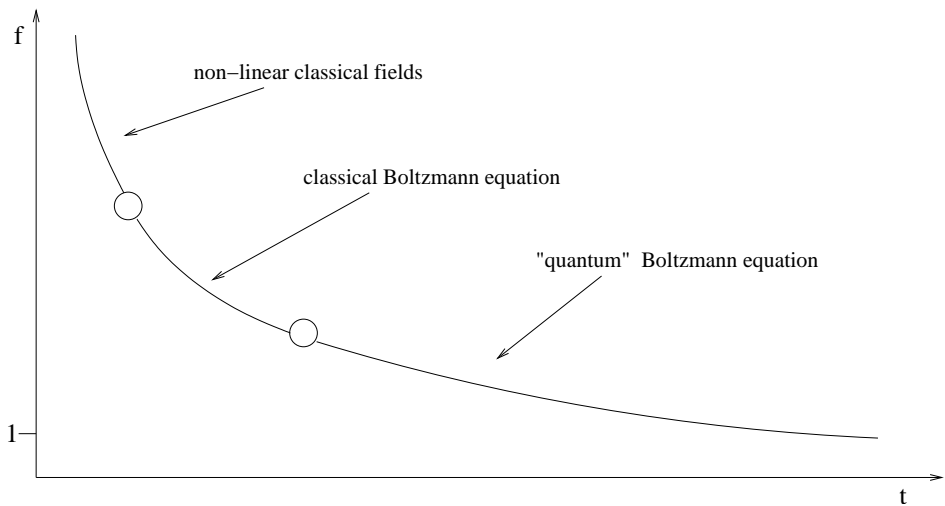,width=12.5cm}
\caption{time evolution of the gluon occupation number}
\end{center}\end{figure}
\noindent However, it is a priori not clear that these transitions are as smooth as the above figure suggests.
In this paper we compare the classical and quantum Boltzmann equations in high energy QCD and show that 
they are equivalent to orders $f^3$ and $f^2$.
The comparison of non-linear classical fields with the classical Boltzmann
equation is an entirely different problem and will not be treated here.

The structure of the paper is as follows: in section 2 we briefly review the main results of Mueller and Son \cite{Mueller:2002gd} where the basis for our considerations is developed within a 
scalar field theory. In section 3 we generalize their strategy to high energy QCD 
and show that one obtains essentially the same results.

We want to emphasize that our argumentations are mostly qualitative. Perhaps surprisingly, we 
need very few computations to derive our results; some simple topological and diagrammatical 
considerations are sufficient. Similar methods can be found, e.g., in \cite{Aarts:1997kp}.

\section{Scalar field theory}

\subsection{Separation of classical and quantum contributions}

Our starting point is the scalar field theory with $\lambda\varphi^4$ interaction given by 
the Lagrangian
\begin{equation}
\mathcal{L}=\frac{1}{2}\partial_\mu\varphi\partial^\mu\varphi
-\frac{1}{2}m^2\varphi^2 -\frac{\lambda}{4!}\varphi^4 \quad .
\end{equation}
Throughout this paper we assume coupling constants to be small and perform the
relevant computations in first order perturbation theory.

Our ultimate interest will be to gain deeper insight into the time evolution of heavy ion 
collisions. So we have to generalize our theory to finite temperature. It is convenient to apply 
the closed time path formalism (CTP) which leads to a doubling of the field
variables (\cite{Schwinger:1960qe}, \cite{Keldysh:1964ud}): $\varphi$ $\rightarrow$ $\Phi_-$,$\Phi_+$.

The Lagrangian then reads
\begin{equation}
\begin{split}
\mathcal{L}_{CTP}&=\frac{1}{2}\partial_\mu\Phi_-\partial^\mu\Phi_- -\frac{1}{2}m^2\Phi_-^2 -\frac{\lambda}{4!}\Phi_-^4\\ 
&\quad -(\frac{1}{2}\partial_\mu\Phi_+\partial^\mu\Phi_+ -\frac{1}{2}m^2\Phi_+^2 -\frac{\lambda}{4!}\Phi_+^4)\quad .
\end{split}
\end{equation}
We now perform a change of the field variables in order to distinguish the classical 
field (denoted $\Phi$) and quantum fluctuations (denoted $\Pi$)
\begin{equation}
\Phi=\frac{1}{2}(\Phi_-+\Phi_+)
\hspace*{15mm}
\Pi=\Phi_--\Phi_+\quad ,
\end{equation}
\begin{equation}
\Phi_-=\Phi+\frac{1}{2}\Pi
\hspace*{15mm}
\Phi_+=\Phi-\frac{1}{2}\Pi\quad .
\end{equation}
As we are interested in systems where both $\Phi_+$ and $\Phi_-$ are 
large, the above interpretations of $\Phi$ and $\Pi$ are already at least 
qualitatively justified.

In terms of these new fields the Lagrangian becomes
\begin{equation}
\mathcal{L}_{\Phi\Pi}=
\partial_\mu\Phi\partial^\mu\Pi - 
m^2\Phi\Pi - 
\frac{\lambda}{3!}(\Phi^3\Pi + \frac{1}{4}\Phi\Pi^3)\quad .
\end{equation}
Indeed, it can be easily shown now by neglecting 
higher than linear terms in $\Pi$ that $\Phi$ fulfills the classical equation 
of motion (for details see \cite{Mueller:2002gd})
\begin{equation}
(\square + m^2)\Phi + 
\frac{\lambda}{3!}\Phi = 0\quad .
\end{equation}

\noindent From the Lagrangian (5) one can identify the vertices of the
theory (Fig. 2).
\begin{figure}[h]\begin{center}
\epsfig{file=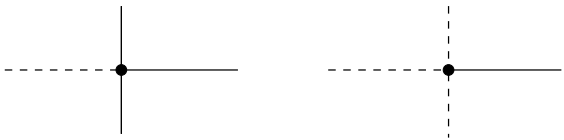,height=1.8cm}
\caption{vertices of the scalar theory}
\end{center}\end{figure}

\noindent The full lines correspond to the classical field $\Phi$ and the dashed
lines to the quantum fluctuations $\Pi$. The diagram on the left will be
called classical vertex, the right one quantum vertex.

The most important quantities for our argumentations will be the free Greens
functions or propagators for the $\Phi$ and $\Pi$ fields, respectively. In a
more rigorous discussion one should adopt the full propagators, but mass
corrections are negligible as long as the coupling constant is small enough,
i.e. $f\lambda \ll 1$. (For details on this point see \cite{Mueller:2002gd}.)

Note that there are also mixed propagators describing the change from $\Phi$ to
$\Pi$ and vice versa.
One obtains

\begin{equation}
G_{\Phi\Phi} = 2\pi\delta(p^2-m^2)(f+\frac{1}{2})\quad ,
\end{equation}
\begin{equation}
G_{\Pi\Phi} = \frac{i}
{p^2-m^2-i\varepsilon p_0}\quad ,
\end{equation}
\begin{equation}
G_{\Phi\Pi} = \frac{i}
{p^2-m^2+i\varepsilon p_0}\quad ,
\end{equation}
\begin{equation}
G_{\Pi\Pi} = 0\quad .
\end{equation}

\noindent The crucial observation here is that only $G_{\Phi\Phi}$
depends on the occupation number f. 

\subsection{Boltzmann equation}

The Boltzmann equation describes the time evolution of the occupation number 
f of a given state as the difference between the scattering of particles 
into and out of this state (gain and loss). These scatterings are included in
the collision term C.
Considering the scalar field theory (1) one may 
express the collision term to lowest order diagrammatically (Fig. 3).
\begin{figure}[h]\begin{center}
\epsfig{file=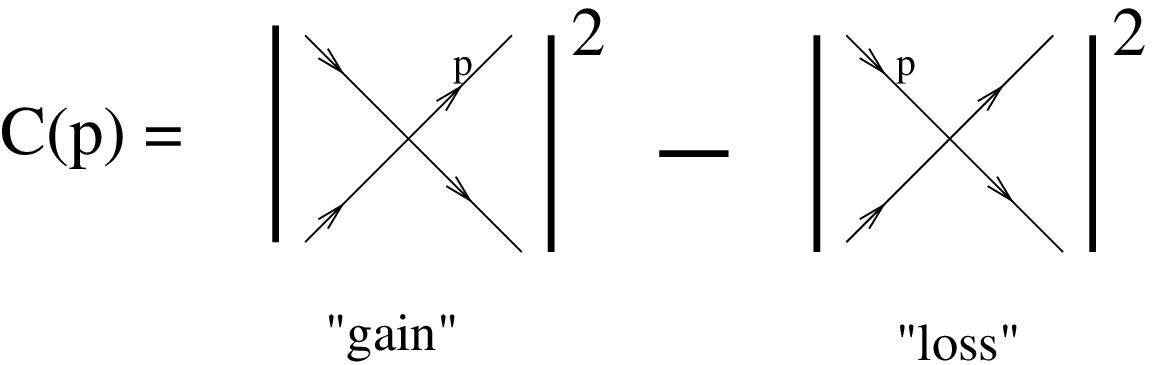,height=1.8cm}
\caption{collision term to lowest order}
\end{center}\end{figure}

\noindent For a detailed analysis of the Boltzmann equation in scalar and gauge field
theories see \cite{Arnold:1998cy}.
For our purposes it is sufficient to examine the topology of the involved 
diagrams and it is easy to see that these are effectively sunset graphs (Fig. 4).
\begin{figure}[h]\begin{center}
\epsfig{file=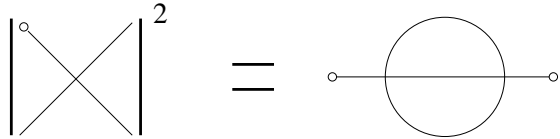,height=1.8cm}
\caption{collision term topology}
\end{center}\end{figure}

\noindent When we express now our theory in terms of $\Phi$ and $\Pi$ we use the corresponding vertices. The classical approximation then consists 
in retaining only collision term diagrams without quantum vertices.

There are three different classical diagrams (Fig. 5).
\begin{figure}[h]\begin{center}
\epsfig{file=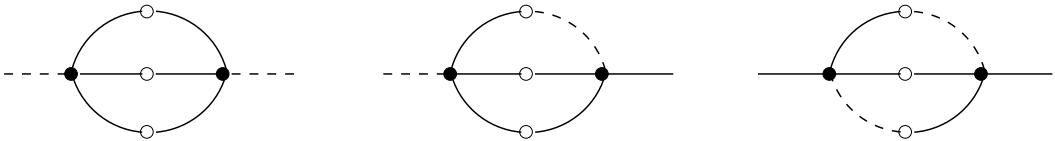,height=1.8cm}
\caption{classical contribution to $C(p)$}
\end{center}\end{figure}

\noindent In Fig. 5 the left diagram is proportional to $(f+\frac{1}{2})^3$ as three propagators 
$G_{\Phi\Phi}$ appear. (Here and in the following empty circles appear in
propagators while full circles stand for vertices.)
Similarly, the other graphs are proportional to $(f+\frac{1}{2})^2$ and
$(f+\frac{1}{2})^1$, respectively.

In the full Boltzmann equation one also has to include the diagrams with one
quantum vertex (Fig. 6).
\begin{figure}[h]\begin{center}
\epsfig{file=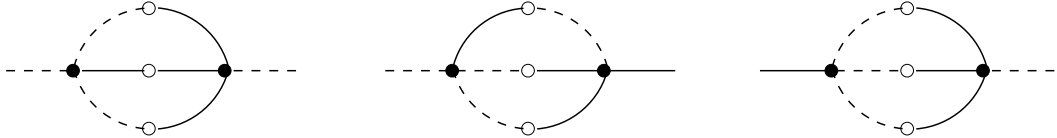,height=1.8cm}
\caption{quantum contribution to $C(p)$}
\end{center}\end{figure}
In Fig.6 the left diagram is proportional to $(f+\frac{1}{2})^1$ while the others
are independent of f.

There are no first order diagrams with two quantum vertices as at least one
propagator $G_{\Pi\Pi} = 0$ would appear.
So we can conclude that classical 
field theory and the Boltzmann equation are equivalent in 
orders $f^3$ and $f^2$ (\cite{Mueller:2002gd}). In the next section we will show how this strategy 
can be generalized to QCD of heavy ion collisions.
\section{High energy QCD}

\subsection{Separation of classical and quantum contributions} 

The results of the previous section may be generalized to QCD by making two 
approximations which are well established in the context of heavy ion 
collisions.

First, we will work within the gluon saturation scenario
(\cite{Gribov:1984tu}, \cite{Blaizot:1987nc}, \cite{McLerran:1993ni}, \cite{McLerran:1994vd}). This means that we may neglect the fermionic degrees of 
freedom in our system. So the QCD Lagrangian simplifies considerably and 
reads in the CTP formalism

\begin{equation}
\mathcal{L}_{CTP} = 
-\frac{1}{4}F_a^{\mu\nu}F_{a\mu\nu}[A_{a\mu}^-] + 
\frac{1}{4}F_a^{\mu\nu}F_{a\mu\nu}[A_{a\mu}^+]\quad .
\end{equation}

\noindent Here the letter `a` is a color index and $A_\mu$ the gluon field.
In analogy to the field transformation in the scalar case we now
define
\begin{equation}
\Phi_{a\mu} = \frac{1}{2}(A_{a\mu}^- + A_{a\mu}^+)
\hspace*{15mm} 
\Pi_{a\mu} = A_{a\mu}^- - A_{a\mu}^+ \quad ,
\end{equation}
\begin{equation}
A_{a\mu}^- =\Phi_{a\mu}+\frac{1}{2}\Pi_{a\mu}
\hspace*{15mm} 
A_{a\mu}^+ =\Phi_{a\mu}-\frac{1}{2}\Pi_{a\mu}\quad .
\end{equation}

\noindent This leads to the following expression for $F_{\mu\nu}^{a+}$
\begin{equation}
\begin{split}
F_{\mu\nu}^{a+}&=\partial_\mu A_\nu^{a+} - \partial_\nu A_\mu^{a+} + gf^{abc}A_\mu^{b+}A_\nu^{c+}\\
&=\partial_\mu \Phi_\nu^{a} - \partial_\nu \Phi_\mu^{a} + gf^{abc}\Phi_\mu^{b}\Phi_\nu^{c}\\
&\quad - \frac{1}{2}(\partial_\mu \Pi_\nu^{a} - \partial_\nu \Pi_\mu^{a} - \frac{1}{2}gf^{abc}\Pi_\mu^{b}\Pi_\nu^{c})\\
&\quad -\frac{1}{2}gf^{abc}(\Phi_\mu^b \Pi_\nu^c + \Pi_\mu^b \Phi_\nu^c)\quad ,
\end{split}
\end{equation}
\noindent and similarly for $F_{\mu\nu}^{a-}$.

Next we express the Lagrangian (11) in terms of the fields $\Phi_\mu$ and 
$\Pi_\mu$. Neglecting higher 
than linear terms in the quantum fluctuations $\Pi_\mu$ one obtains

\begin{equation}
\mathcal{L}_\text{linear} = 
(D_\mu^{ab} F_b^{\mu\nu} [\Phi])\Pi_\nu^a 
\end{equation}
with

\begin{equation}
D_\mu^{ab} \equiv \partial_\mu \delta^{ab} - gf^{abc}\Phi_\mu^c \quad ,
\end{equation}
\begin{equation}
F_b^{\mu\nu} [\Phi] \equiv 
\partial^\mu \Phi^\nu_{b} - 
\partial^\nu \Phi^\mu_{b} + 
gf^{bcd}\Phi^\mu_{c}\Phi^\nu_{d} \quad .
\end{equation}

\noindent Thus, $\Phi_\mu$ fulfills the classical equation of motion

\begin{equation}
D_\mu^{ab} F_b^{\mu\nu} [\Phi] = 0 \quad .
\end{equation}

\noindent So our interpretation of $\Phi_\mu$ as classical field is justified in 
complete analogy to the scalar case.

From the equations (15) to (17) one also sees that there is only one
classically allowed vertex in first order in the coupling constant g (Fig. 7). 

\begin{figure}[h]\begin{center}
\epsfig{file=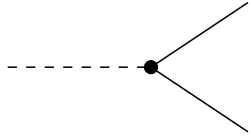,height=1.8cm}
\caption{classical vertex in QCD}
\end{center}\end{figure}
\noindent Here full lines denote classical fields and dashed lines quantum 
corrections like in the previous section. 

In order to obtain the quantum 
couplings one has to take into account the terms nonlinear in $\Pi_\mu$. 
The corresponding Lagrangian may be easily computed and reads

\begin{equation}
\mathcal{L}_{nonlin} = 
-\frac{1}{4}gf_{abc}\Pi^{b\mu}\Pi^{c\nu}
\{\frac{1}{2}(\partial_\mu\Pi_\nu^a - 
\partial_\nu\Pi_\mu^a) + gf^{ade}\Phi_\mu^d\Pi_\nu^e\}\quad .
\end{equation}

\noindent So we have one first order quantum coupling (Fig. 8).

\begin{figure}[h]\begin{center}
\epsfig{file=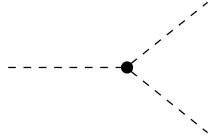,height=1.8cm}
\caption{quantum vertex in QCD}
\end{center}\end{figure}

\subsection{The collision term}

Our second approximation leads to a simple topology of the collision term C. 
As fermions are neglected the only contributions to C come from gluon 
scattering. We now assume that the t-channel dominates, where t is 
the Mandelstam variable (see, e.g., \cite{Cutler:1977qm}).
(In fact, our only assumption is to work in the high energy limit where both approximations are valid.)

The t-channel gluon scattering is shown diagrammatically in Fig. 9.

\begin{figure}[h]\begin{center}
\epsfig{file=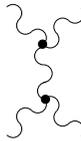,height=1.8cm,}
\caption{t-channel gluon scattering}
\end{center}\end{figure}
\noindent This makes the relevant collision term topology quite simple
as can be seen in Fig. 10 (note that in contrast to the previous section only three-field vertices are
present).

\begin{figure}[h]\begin{center}
\epsfig{file=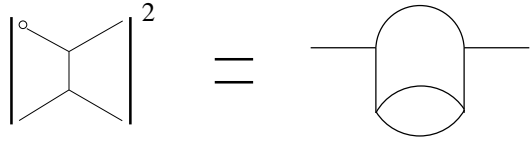,height=1.8cm}
\caption{collision term topology in QCD}
\end{center}\end{figure}
As the main features of the propagators do not change, especially 
$G_{\Phi\Phi} \propto (f+\frac{1}{2})$ and $G_{\Pi\Pi} = 0$, we are now 
again ready to compare classical field theory with the full Boltzmann 
equation. As before the classical approximation consists in neglecting
diagrams with quantum vertices which are only taken into account in the full collision 
term. 

Let us consider for example the  diagram in Fig. 11 that clearly has the
required topology.

\begin{figure}[h]\begin{center}
\epsfig{file=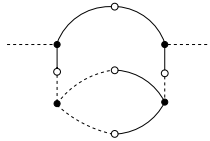,height=1.8cm}
\caption{quantum contribution to $C(p)$}
\end{center}\end{figure}
\noindent It contains one quantum vertex so it is not included in the
classical approximation. The contribution from this graph to the collision term is proportional to $(f+\frac{1}{2})$ as one propagator $G_{\Phi\Phi}$ appears.
Similarly, the diagram in Fig. 12 is classical and proportional to
$(f+\frac{1}{2})^3$.
\begin{figure}[h]\begin{center}
\epsfig{file=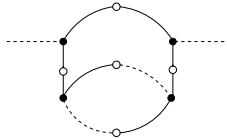,height=1.8cm}
\caption{classical contribution to $C(p)$}
\end{center}\end{figure}
It is easy to check that these examples give already the highest order 
contributions to the classical approximation and the quantum corrections, 
respectively. 

We conclude that the results of the scalar case are indeed valid in 
high energy QCD: The classical field approximation and the quantum  Boltzmann 
equation match in orders $f^3$ and $f^2$, i.e. in leading orders of the gluon
occupation number.

\section{Acknowledgments}

I would like to thank Rolf Baier for advice and discussions.\\ 
This work is supported by DFG.


\begin{thebibliography}{00}

\bibitem{Mueller:2002gd}
A.~H.~Mueller and D.~T.~Son,
Phys.\ Lett.\ B {\bf 582} (2004) 279
[arXiv:hep-ph/0212198].

\bibitem{Ludlam:2003rh}
T.~Ludlam and L.~McLerran,
Phys.\ Today {\bf 56N10} (2003) 48.

\bibitem{Gyulassy:2004zy}
M.~Gyulassy and L.~McLerran,
arXiv:nucl-th/0405013.

\bibitem{Mueller:1999pi}
A.~H.~Mueller,
Phys.\ Lett.\ B {\bf 475} (2000) 220
[arXiv:hep-ph/9909388].

\bibitem{Mueller:1999fp}
A.~H.~Mueller,
Nucl.\ Phys.\ B {\bf 572} (2000) 227
[arXiv:hep-ph/9906322].

\bibitem{Aarts:1997kp}
G.~Aarts and J.~Smit,
Nucl.\ Phys.\ B {\bf 511} (1998) 451
[arXiv:hep-ph/9707342].

\bibitem{Schwinger:1960qe}
J.~S.~Schwinger,
J.\ Math.\ Phys.\  {\bf 2} (1961) 407.

\bibitem{Keldysh:1964ud}
L.~V.~Keldysh,
Zh.\ Eksp.\ Teor.\ Fiz.\  {\bf 47} (1964) 1515
[Sov.\ Phys.\ JETP {\bf 20} (1965) 1018].

\bibitem{Arnold:1998cy}
P.~Arnold, D.~T.~Son and L.~G.~Yaffe,
Phys.\ Rev.\ D {\bf 59} (1999) 105020
[arXiv:hep-ph/9810216].

\bibitem{Gribov:1984tu}
L.~V.~Gribov, E.~M.~Levin and M.~G.~Ryskin,
Phys.\ Rept.\  {\bf 100} (1983) 1.

\bibitem{Blaizot:1987nc}
J.~P.~Blaizot and A.~H.~Mueller,
Nucl.\ Phys.\ B {\bf 289} (1987) 847.

\bibitem{McLerran:1993ni}
L.~D.~McLerran and R.~Venugopalan,
Phys.\ Rev.\ D {\bf 49} (1994) 2233
[arXiv:hep-ph/9309289].

\bibitem{McLerran:1994vd}
L.~D.~McLerran and R.~Venugopalan,
Phys.\ Rev.\ D {\bf 50} (1994) 2225
[arXiv:hep-ph/9402335].

\bibitem{Cutler:1977qm}
R.~Cutler and D.~W.~Sivers,
Phys.\ Rev.\ D {\bf 17} (1978) 196.


\end{thebibliography}
\end{document}